\documentclass[twocolumn,amsmath,amssymb,prl,showpacs]{revtex4}

\def\Re{\hbox{Re}\,}
\def\Im{\hbox{Im}\,}
\def\lsim{\mathrel{\rlap{\lower3pt\hbox{$\sim$}}
    \raise2pt\hbox{$<$}}}
\newcommand{\rf}[1]{(\ref{#1})}

\begin{document}

\title{Vacuum \v{C}erenkov radiation}

\author{Ralf Lehnert}
%\email[Electronic mail: ]{rlehnert@ualg.pt}%

\author{Robertus Potting}%
%\email[Electronic mail: ]{rpotting@ualg.pt}%

\affiliation{CENTRA, Departamento de F\'{\i}sica,
Universidade do Algarve, 
8000-117 Faro, Portugal}

\date{June 10, 2004}

\begin{abstract}
Within the classical Maxwell--Chern--Simons limit
of the Standard-Model Extension (SME), 
the emission of light 
by uniformly moving charges 
is studied 
confirming the possibility of a \v{C}erenkov-type effect. 
In this context, 
the exact radiation rate 
for charged magnetic point dipoles is determined 
and found in agreement with a phase-space estimate 
under certain assumptions.
\end{abstract}


\maketitle

The \v{C}erenkov effect---emission of radiation 
from charges moving at or above
the phase speed of light---is
experimentally and theoretically 
well established 
in conventional macroscopic media \cite{cer1}, 
and its importance 
for modern particle detectors 
in high-energy and cosmic-ray physics 
can hardly be overstated. 
The reasons behind 
the recent revival of interest 
in this subject are twofold.

First, 
conventional-physics investigations, 
such as observations at CERN 
involving lead ions \cite{cern} 
and experiments in exotic condensed-matter systems \cite{cms}, 
have found unexpected features of the \v{C}erenkov effect. 
They include 
nonstandard kinematical radiation conditions, 
backward photon emission, 
and backward-pointing \v{C}erenkov cones. 
Some of these issues have been studied 
theoretically \cite{certheo}. 

Second,
many candidate models 
underlying established physics 
predict Lorentz-breaking vacua \cite{cpt01}, 
in which modified light speeds, 
for instance, 
offer the possibility of Lorentz tests 
via a \v{C}erenkov-type mechanism 
called ``vacuum \v{C}erenkov radiation.''
At presently attainable energies, 
this and other Lorentz-breaking effects 
are described by
the Standard-Model Extension (SME) \cite{sme}.
Candidate underlying models
include strings \cite{kps},
spacetime foam \cite{lqg},
noncommutative geometry \cite{ncft},
varying scalars \cite{vc},
random-dynamics models \cite{rd},
multiverses \cite{mv},
and brane worlds \cite{bws}.
Numerous analyses of Lorentz breaking in
mesons,
baryons,
electrons,
photons,
muons,
neutrinos,
and the Higgs sector
have been performed
within the SME \cite{cpt01}.
Although of substantial importance
for Lorentz-violation studies \cite{vcr,bws},
a detailed investigation of vacuum \v{C}erenkov radiation
is currently still lacking.

The present work
is primarily intended to fill this gap. 
However, 
we expect our analysis 
to remain applicable also 
for conventional macroscopic media. 
In particular, 
our study provides a new conceptual perspective 
on \v{C}erenkov radiation 
augmenting the usual physics picture: 
our fully relativistic Lagrangian 
allows us to work in the charge's rest frame, 
where fields typically behave like 
$r^{-1}\exp(-\sqrt{p\hspace{-1pt}\cdot\hspace{-1pt}p}\,r)$
at large distances $r$ from the source.
Here, $p^{\mu}$ 
satisfies the plane-wave dispersion relation, 
and the metric has signature $-2$. 
Conventional massive fields $p\hspace{-1pt}\cdot\hspace{-1pt}p=m^2$
lead to the Yukawa potential $r^{-1}\exp(-mr)$,
and the massless limit gives the standard $r^{-1}$ behavior. 
In these cases, 
the energy--momentum tensor
vanishes rapidly for $r\to\infty$ 
precluding energy--momentum flux to infinity.
However, 
Lorentz-violating vacua and macroscopic media 
permit $p\hspace{-1pt}\cdot\hspace{-1pt}p<0$
resulting in oscillatory far fields
and thus the possibility of radiation. 
We show
that this idea leads
to the standard radiation conditions
and facilitates the determination of
the exact emission rate for charged magnetic dipoles
within our dispersive and anisotropic model. 
To our knowledge, 
this is in many respects the first such result.

The present analysis employs
the classical Maxwell--Chern--Simons limit
of the SME given by the Lagrangian
\begin{equation}
{\cal L}_{\rm MCS} = 
-\frac{1}{4} F_{\mu\nu}F^{\mu\nu}
+k_{\mu}A_{\nu}\tilde{F}^{\mu\nu}
-A_{\mu}j^{\mu}. 
\label{lagr}
\end{equation}
Here $F_{\mu\nu}=\partial_{\mu}A_{\nu}-\partial_{\nu}A_{\mu}$
denotes the conventional electromagnetic field-strength tensor 
and $\tilde{F}^{\mu\nu}={\textstyle{1\over 2}}
\varepsilon^{\mu\nu\rho\sigma}F_{\rho\sigma}$ its dual,
as usual.
We have included an external source 
$j^{\mu}=(\varrho,
\hspace{1.5pt}\vec{\hspace{-1.5pt}\textit{\j}}\hspace{1pt}\hspace{1pt})$
and adopted natural units 
$c\hspace{-1pt} =\hspace{-1pt} \hbar\hspace{-1pt} =\hspace{-1pt} 1$. 
The spacetime-constant nondynamical $k^{\mu}= (k^0,\vec{k}\hspace{1pt})$
violates Lorentz, PT, and CPT invariance \cite{sme,rl03}.
Although tightly constrained observationally \cite{cfj}, 
this model has been studied extensively in the literature \cite{sme,cfj,mcs}. 

The potential $A^{\mu}$
obeys the equation of motion 
\begin{equation}
\left(\Box \eta^{\mu\nu}-\partial^{\mu} \partial^{\nu}
-2\varepsilon^{\mu\nu\rho\sigma}k_{\rho}\partial_{\sigma}\right)A_{\nu}
=j^{\mu} .
\label{oddeom}
\end{equation}
Paralleling the conventional case, 
current conservation $\partial_{\mu}j^{\mu}=0$ 
is required for consistency.
Equation \rf{oddeom}
gives the following modified Coulomb
and Amp\`ere laws: 
\begin{eqnarray}
\vec{\nabla}\!\cdot\!\vec{E}-2\vec{k}\!\hspace{0.8pt}\cdot\!\vec{B} & 
= & \varrho ,\nonumber\\
-\hspace{1.5pt}\dot{\hspace{-1.5pt}\vec{E}}+\vec{\nabla}\!\times\!\vec{B}
-2k_0\vec{B}+2\vec{k}\!\times\!\vec{E} & = & \hspace{1.5pt}\vec{\hspace{-1.5pt}\textit{\j}}\hspace{1pt} .
\label{oddmax}
\end{eqnarray}
The field-potential relationship 
is unaltered, 
so that the homogeneous Maxwell equations 
remain unchanged.  
Gauge invariance of physics 
is evident from Eqs.\ \rf{oddmax}, 
and any of the usual conditions on $A^{\mu}$, 
like Lorentz or Coulomb gauge,  
can be imposed \cite{sme}. 
Equation \rf{oddeom} implies 
that for $j^{\mu}\neq 0$ the energy--momentum tensor
\begin{equation} 
\Theta^{\mu\nu}=-F^{\mu\alpha}F^{\nu}{}_{\!\alpha} 
+\frac{1}{4}\eta^{\mu\nu}F^{\alpha\beta}F_{\alpha\beta} 
-k^{\nu}\tilde{F}^{\mu\alpha}A_{\alpha} 
\label{emtensor} 
\end{equation}
is in general not conserved, as expected: 
\begin{equation}
\partial_{\mu}\Theta^{\mu\nu}=j_{\mu}F^{\mu\nu}. 
\label{emcurrent} 
\end{equation}
Although $\Theta^{\mu\nu}$ depends on $A^{\mu}$, 
the physical 4-momentum remains gauge invariant \cite{cfj}. 

Up to homogeneous solutions, 
Eq.\ \rf{oddeom} is solved by 
\begin{equation}
A^{\mu}(x)=\int\limits_{C_\omega}\hspace{-1.0pt}
\frac{d^4 p}{(2\pi)^4}\;G^{\mu\nu}
\textit{\^{\j}}\hspace{.5pt}
_{\nu}
\exp (-ip\!\hspace{0.8pt} \cdot \! x),
\label{oddsln}
\end{equation}
where 
\begin{equation}
G^{\mu\nu} \equiv -\frac{p^2 \eta^{\mu\nu}
+2i\varepsilon^{\mu\nu\rho\sigma}k_{\rho}p_{\sigma}
+4k^{\mu}k^{\nu}}{p^4+4p^2k^2-4(p\!\hspace{0.8pt}\cdot\! k)^2}. 
\label{oddgreen}
\end{equation}
Here, $x^{\mu}=(t,\vec{r}\hspace{1pt})$
is the spacetime-position vector 
and $p^{\mu}=(\omega,\vec{p}\hspace{1pt})$ the Fourier-space wave vector.
A caret denotes the four-dimensional Fourier transform. 
The poles of the integrand in Eq.\ \rf{oddsln} 
give the dispersion relation
\begin{equation}
p^4+4p^2k^2-4(p\!\hspace{0.8pt}\cdot\! k)^2=0 .
\label{odddisp}
\end{equation} 
To ensure causal propagation, 
the $\omega$-integration contour $C_{\omega}$ 
must pass above all poles on the real-$\omega$ axis, 
as usual. 
This is best implemented 
by replacing $\omega\rightarrow\omega+i\varepsilon$ in each $\omega$ 
in the denominator
of the integrand in Eq.\ \rf{oddsln}. 
The infinitesimal positive parameter $\varepsilon$
is taken to approach zero after the integration. 
Note, 
however, 
that for timelike $k^{\mu}$ 
poles on the imaginary-$\omega$ axis 
occur, 
so that causality is violated \cite{cfj,mcs}. 
In what follows,
we therefore
focus on the spacelike- and lightlike-$k^{\mu}$ cases.

The current distribution 
describing the particle 
should be time independent 
in the particle's rest frame, 
so that 
$\textit{\^{\j}}\hspace{1pt}
^{\mu}(p^{\mu})
=2\pi\delta(\omega)\,\textit{\~{\j}}\hspace{1pt}
^{\mu}(\vec{p}\hspace{1pt})$,
where the tilde denotes the three-dimensional Fourier transform.
Then, 
Eq.\ \rf{oddsln} takes the form 
\begin{equation}
A^{\mu}(\vec{r}\hspace{1pt}) = 
\int \hspace{-1.0pt}\frac{d^3 \vec{p}}{(2\pi)^3}\;
\frac{N^{\mu\nu}(\vec{p}\hspace{1pt})\,
\textit{\~{\j}}\hspace{.5pt}
_{\nu}(\vec{p}\hspace{1pt})
\exp(i\vec{p}\!\hspace{0.8pt}\cdot\!\vec{r}\hspace{1pt})}
{\vec{p}^{\,4}-4\vec{p}^{\,2}k^2
-4(\vec{p}\!\hspace{0.8pt}\cdot\!\vec{k}-i\varepsilon k_0)^2},
\label{Asoln}
\end{equation}
where $N^{\mu\nu}(\vec{p}\hspace{1pt})\equiv\vec{p}^{\,2}\eta^{\mu\nu}
-2i\varepsilon^{\mu\nu\rho s}k_{\rho}p_s
-4k^{\mu}k^{\nu}$, 
and Latin indices run from 1 to 3. 
Evaluation of the $|\vec{p}\hspace{1pt}|$-type integral
with complex-analysis methods 
gives certain residues of the integrand 
in the complex $|\vec{p}\hspace{1pt}|$ plane,
which typically contain the factor
$\exp (i\vec{p}_0\!\cdot\!\hspace{0.8pt}\vec{r}\hspace{1pt})$. 
Here, $\vec{p}_0$ denotes the location of a pole. 
The remaining angular integrations
correspond to averaging the residues
over all directions, 
so that the qualitative behavior of the integral \rf{Asoln} 
is determined by the residues. 
In particular, 
$A^{\mu}$ decreases exponentially with increasing $r$ 
for $\Im(\vec{p}_0)\neq \vec{0}$, 
while $\Re(\vec{p}_0)\neq \vec{0}$ leads to oscillations with distance. 
As mentioned earlier, 
energy transport to spatial infinity requires 
non-decaying oscillatory fields. 
\textit{Thus, 
one expects vacuum \v{C}erenkov radiation only
when there are real $p^{\mu}=(0,\vec{p}\hspace{1pt})$
satisfying the plane-wave dispersion relation
in the source's rest frame.}
In a general inertial frame,
this condition reads
$p'^{\mu}=(\vec\beta\hspace{-1pt}\cdot\hspace{-1pt}\vec{p}\hspace{1pt}'\!,
\vec{p}\hspace{1pt}')$,
where $\vec\beta$ denotes the velocity of the particle.
This is seen to be equivalent
to the conventional phase-speed condition
$c\hspace{0.8pt}'_{ph}=|\omega'\hspace{-1.5pt}|/
|\vec{p}\hspace{1pt}'\hspace{-1.5pt}|\le |\vec\beta|$.
Note 
that spacelike plane-wave vectors 
are not necessarily associated with 
positivity and stability problems \cite{vc}. 

The usual method for calculating the radiation rate---extraction 
of the $r^{-2}$ piece 
of the modified Poynting vector 
and integration over a spherical surface---is challenging 
because the determination of the far fields 
is hampered 
by the complexity of the integral \rf{Asoln}. 
For further progress, 
an ansatz for the current $j^{\mu}=J^{\mu}$ 
describing the particle 
is advantageous. 
The most general form of $J^{\mu}(x)$ 
consistent with current conservation 
and the presumed time independence in the
particle's
rest frame is
\begin{equation} 
J^{\mu}(\vec{r}\hspace{1pt})=\big( \rho(\vec{r}\hspace{1pt}),
\vec{\nabla}\!\times\!\vec{f}(\vec{r}\hspace{1pt})\big),
\label{currentansatz} 
\end{equation}
where $\rho(\vec{r}\hspace{1pt})$ is the source's charge density 
and $\vec{f}(\vec{r}\hspace{1pt})$ is an arbitrary vector field. 
Moreover, 
we require both $\rho(\vec{r}\hspace{1pt})$ 
and $\vec{f}(\vec{r}\hspace{1pt})$ 
to vanish rapidly outside the 
finite volume $V_0$ 
associated with the particle. 
We can therefore drop 
various boundary terms in the subsequent manipulations,
if the integration volume $V$ is chosen large enough.

Spatial integration of Eq.\ \rf{emcurrent} yields 
\begin{equation} 
\int\limits_{\sigma}\hspace{-1.0pt} d\sigma^{l} \, \Theta_{l\nu}
= \int\limits_{V}\hspace{-1.0pt}d^3\vec{r} \; J^{\mu}F_{\mu\nu}
- \frac{\partial}{\partial t} \int\limits_{V}\hspace{-1.0pt}d^3\vec{r} \; \Theta_{0\nu},
\label{encons} 
\end{equation} 
where $\sigma$ is the boundary of $V$,
and $d\sigma^{l}$ the corresponding surface element 
with outward orientation. 
The energy--momentum flux
${}\hspace{2pt}\dot{\hspace{-2pt}P}_{\hspace{-1pt}\nu}\equiv\int_{\sigma} d\sigma^{l} \, \Theta_{l\nu}$
through the surface $\sigma$ 
is therefore 
caused by the source $J^{\mu}(x)$ in the enclosed volume $V$ 
and the decrease in the field's 
4-momentum in $V$, as usual.
Using the Maxwell equation $\vec\nabla\times \vec E=0$
and the zeroth component of Eq.\ \rf{encons}
one obtains
the following expression 
for the radiated energy: 
\begin{equation}
\int\limits_{\sigma} \hspace{-1.0pt}d\vec{\sigma}\cdot\vec{S}
= -\int\limits_{\sigma} \hspace{-1.0pt}d\vec{\sigma}\cdot(\vec{f}\!\times\!\vec{E}),
\label{poy}
\end{equation}
where a modified Poynting vector
$\Theta_{l0}\equiv S_{l}=-S^{l}$ has been defined.
Since $\vec{f}$ goes to zero 
on the boundary of a large volume, 
the energy flux to infinity vanishes. 
This feature is model independent:
in the particle's rest frame, 
$J^{\mu}$ is spatially localized, 
and time-translation invariance holds. 
The resulting energy conservation 
implies zero energy flux 
through any closed surface. 
{\it Thus, the net radiated energy vanishes 
in the rest frame of a localized static source.}
Note that any nonzero 4-momentum 
radiated by such a source 
is therefore necessarily spacelike.

The 3-momentum emission rate
is obtained similarly:
\begin{equation}
{}\hspace{1.5pt}\dot{\hspace{-1.5pt}\vec{P}}
=\int\limits_{V}\hspace{-1.0pt}d^3\vec{r} \;
J_{\mu}\,\vec{\nabla}\hspace{-1.0pt} A^{\mu} .
\label{3mom}
\end{equation}
Employing Eq.\ \rf{Asoln}
and the Fourier expansion of $J^{\mu}$ yields
\begin{equation}
{}\hspace{1.5pt}\dot{\hspace{-1.5pt}\vec{P}}
=i\hspace{-1.0pt}\int\limits\hspace{-1.0pt}\frac{d^3 \vec{p}}{(2\pi)^3}
\;\frac{\tilde{J}^{\mu}(-\vec{p}\hspace{1pt})N_{\mu\nu}(\vec{p}\hspace{1pt})\tilde{J}^{\nu}(\vec{p}\hspace{1pt})}
{\vec{p}^{\,4}-4\vec{p}^{\,2}k^2-4(\vec{p}\!\hspace{0.8pt}\cdot\!\vec{k}-i\varepsilon k_0)^2}\;\vec{p}
\label{restrate}
\end{equation}
in the limit $V\rightarrow\infty$. 
Note that the integrand is odd in $\vec{p}$, 
so that $\hspace{1.5pt}\dot{\hspace{-1.5pt}\vec{P}}$ vanishes 
if singularities are absent.
However, 
this symmetry argument fails 
for integrands with poles at real $\vec{p}=\vec{p}_0$,
consistent with our radiation condition.
The dispersion relation \rf{odddisp}
indeed admits such solutions 
opening a doorway 
for nonzero 3-momentum emission.

We now focus
on the current distribution
$\rho=q\,\delta(\vec{r}\hspace{1pt})$ and
$\vec J=-\vec\mu\!\,\times\!\vec{\nabla}\delta(\vec{r}\hspace{1pt})$,
which describes a point-like charge $q$
with magnetic dipole moment $\vec\mu$.
The use of a suitable regulation of the delta-function
then permits a closed-form evaluation
of the integral in Eq.\ \rf{restrate}.
This gives the exact rest-frame rate
of 3-momentum radiation for 
a charged magnetic point dipole:
\begin{widetext}
\begin{equation} 
{}\hspace{1.5pt}\dot{\hspace{-1.5pt}\vec{P}}
=-\frac{{\rm sgn}(k_0)}{12\pi}\; 
\frac{k_0^5}{|\vec{k}|^5}
\left\{
\left[3q^2 \vec{k}^{\,2}/k_0^2 +6q\,\vec{k}\!\hspace{0.8pt}\cdot\!\vec{\mu}
-\vec{\mu}^{\,2}k_0^2
+5(\vec{k}\!\hspace{0.8pt}\cdot\!\vec{\mu})^2k_0^2/\vec{k}^{\,2}
+10(\vec{k}\!\hspace{0.8pt}\times\!\vec{\mu})^2\right]k_0\vec{k}
-2\left[q\vec{k}^{\,2}/k_0^2+\vec{k}\!\hspace{0.8pt}\cdot\!\vec{\mu}\right]k_0^3\,\vec{\mu}\right\}.
\label{drate} 
\end{equation} 
\end{widetext}
A nonzero flux 
in the above static case 
might appear counter-intuitive.
However, similar situations 
arise in conventional physics as well. 
For instance,
constant non-parallel $\vec{E}$ and $\vec{B}$ fields
are associated
with a finite Poynting flux $\vec{S}=\vec{E}\!\times\!\vec{B}$.
Although suppressed by four powers of $k^{\mu}$, 
the rate \rf{drate} remains nonvanishing 
in the zero-charge limit $q\to 0$.
Ordinary refractive indices 
typically require a minimal speed of the charge 
for the emission of \v{C}erenkov light.
This holds no longer true in the present context, 
as can be seen in the case 
for lightlike $k^{\mu}$ and $\vec{\mu}=\vec{0}$. 
{\it Thus, vacuum \v{C}erenkov radiation
need not necessarily be a threshold effect.}

The radiation rate in the laboratory frame 
is often more useful. 
To avoid unwieldy expressions, 
we consider the special case of vanishing $\vec{\mu}$
and spacelike $k^{\mu}$.
We further choose the laboratory such that
$k_0'=0$ and $\vec{k}'\neq\vec{0}$.
Then, suppressing the primes,
Eq.\ \rf{drate} becomes
\begin{equation}
{}\hspace{2pt}\dot{\hspace{-2pt}P}_{\hspace{-1pt}\mu}
=\frac{q^2}{4\pi}\;
\frac{\gamma^3(\vec{\beta}\!\hspace{0.8pt}\cdot\!\vec{k}\hspace{0.8pt})^4}
{\vec{k}^{\,2}+\gamma^2(\vec{\beta}\!\hspace{0.8pt}\cdot\!\vec{k}\hspace{0.8pt})^2}\;
K_{\mu} ,
\label{labpcrate}
\end{equation}
where
\begin{equation}
K^{\mu}\equiv\frac{{\rm sgn}(\vec{\beta}\!\hspace{0.8pt}\cdot\!\vec{k}
\hspace{0.8pt})}
{\sqrt{\vec{k}^{\,2}+\gamma^2(\vec{\beta}\!\hspace{0.8pt}\cdot\!\vec{k}
\hspace{0.8pt})^2}}\;
{ \gamma^2(\vec{\beta}\!\hspace{0.8pt}\cdot\!\vec{k}\hspace{0.8pt}) \choose
\;\vec{k}+\gamma^2(\vec{\beta}\!\hspace{0.8pt}\cdot\!\vec{k}\hspace{0.8pt})\vec{\beta}\; }.
\label{dir}
\end{equation}
Here, $\vec{\beta}$ is the 3-velocity of the charge
in the laboratory
and $\gamma=(1-\vec{\beta}^2)^{-1/2}$. 
The over-dot now denotes differentiation 
with respect to laboratory time.
For particle 3-velocities perpendicular to $\vec{k}$,
radiation is absent.
In all other cases,
both $\dot P^0 $
and the projection of ${}\hspace{1.5pt}\dot{\hspace{-1.5pt}\vec{P}}$ 
onto $\vec{\beta}$
are positive,
so that conventional charges
are decelerated
in our laboratory frame.
Note, 
however, 
that 
as a consequence of the anisotropic vacuum, 
the net emitted 3-momentum 
is typically not aligned 
with the charge's velocity. 
The back-reaction 
of the radiation 
on the charge 
will then in general 
lead to a curved trajectory for the particle. 
Regardless of anisotropies, 
4-momentum loss implies nongeodesic motion. 
{\it Vacuum \v{C}erenkov radiation
is therefore always associated 
with Equivalence-Principle violations.} 

Most cosmic-ray analyses of Lorentz violation
are based on purely kinematical models, 
so that it is interesting to study, 
whether a modified dispersion relation by itself 
permits a sensible estimate for the \v{C}erenkov rate. 
In quantum theory, 
the \v{C}erenkov effect corresponds to 
the decay of a charge $P_a$
into itself $P_b$ 
and a photon $P_c$. 
In the center-of-mass frame, 
the rate for this process obeys
\begin{equation}
d\Gamma=\frac{|{\cal M}_{a\rightarrow b,c}|^2}{2m}\,(2\pi)^4\,
\delta^{(4)}(p_a^{\mu}-p_b^{\mu}-p_c^{\mu})\,d\Pi_b\,d\Pi_c,
\label{gendecay}
\end{equation}
where the transition amplitude
${\cal M}_{a\rightarrow b,c}$ 
contains information about the dynamics of the decay.
The remaining factors describe the kinematics of the process.
They include phase-space elements $d\Pi_{s}$
and various 4-momenta $p_s^{\mu}=(E_s,\vec{p}_s)$,
where $s\in\{a,b,c\}$
refers to the corresponding particle.
In what follows,
we consider a conventional charge $q$ with $p_a^2=p_b^2=m^2$.
To facilitate a transparent comparison
with the classical result \rf{drate},
we further assume photon 4-momenta
$p_c^{\mu}$
obeying the dispersion relation \rf{odddisp},
select a lightlike $k^{\mu}$ parameter,
and take the static-source limit $m\to\infty$.

An order-of-magnitude estimate 
for the transition amplitude is 
$|{\cal M}_{a\rightarrow b,c}|^2\sim q^2m^2$ \cite{kauf}, 
where the spinor normalization 
implicit in Eq.\ \rf{gendecay} 
has been used. 
The phase-space element $d\Pi_b$
is determined by the conventional relation
$2E_b(\vec{p}\hspace{1pt})\,d\Pi_b=(2\pi)^{-3}d^3\vec{p}$.
The construction
of the invariant phase-space element $d\Pi_c$
for the photon requires more care
due to the presence of Lorentz breaking.
Coordinate independence requires
\begin{equation}
d\Pi_c=\frac{d^3\vec{p}_c}{(2\pi)^32|\vec{p}_c+{\rm sgn}(k^0)\vec{k}|}
\label{psel}
\end{equation}
for the
positive-energy,
spacelike roots of the dispersion relation \rf{odddisp}.
Noting
that $d\hspace{2pt}\dot{\hspace{-2pt}P}_{\hspace{-1pt}\mu}=p_{\mu}\,d\Gamma$,
our above considerations lead to
\begin{equation}
{}\hspace{1.5pt}\dot{\hspace{-1.5pt}\vec{P}}
\sim
-\frac{q^2}{8\pi}k^0\vec k
\label{momdecay}
\end{equation}
as a rough estimate
for the net radiated momentum per time
in the charge's rest frame.
Comparison with Eq.\ \rf{drate}
supports the validity our phase-space result \rf{momdecay}.
{\it We conclude
that in the context of vacuum \v{C}erenkov radiation
phase-space considerations
can provide useful estimates for momentum-emission rates.}

Experimental studies
employing the \v{C}erenkov effect
in the Maxwell--Chern--Simons model
are unlikely to improve
the existing tight bound of
${\cal O} (k^{\mu})\lsim 10^{-42}\text{ GeV}$ \cite{cfj}.
However, Eq.\ \rf{labpcrate}
identifies an average alignment
of charged-matter velocities
in the plane perpendicular to $\vec{k}$
as a potential signature
in a cosmological context.
This effect might
have been
enhanced
before electroweak symmetry breaking.
First,
radiation is not yet decoupled from the matter
resulting in a large number of free charges
that can be affected.
Second,
lightlike 4-momenta of massless charged matter imply
that all wave frequencies
can contribute to vacuum \v{C}erenkov radiation
\cite{footnote}.
We also note
that our general philosophy and methods
are applicable in other Lorentz-violating situations.
For instance,
some components of the dimensionless $(k_{F})^{\mu\nu\rho\sigma}$ parameter
in the SME
are currently only bounded at the $10^{-9}$ level \cite{photonexpt}.
Moreover,
the rate might be less suppressed in this case
offering the possibility of improved constraints
via vacuum \v{C}erenkov radiation.

In conclusion, 
a generic conceptual picture of the \v{C}erenkov effect, 
which complements the conventional one, 
has been developed 
and illustrated explicitly 
in the Maxwell--Chern--Simons model. 
This physical picture offers 
an interesting avenue 
for further insight into various aspects 
of \v{C}erenkov radiation
in general Lorentz-breaking vacua and macroscopic media. 
It paves the way  
for additional studies in a quantum context 
and provides a solid foundation for phenomenological explorations 
of Lorentz violation via the vacuum \v{C}erenkov effect.

\acknowledgments
We thank Frans Klinkhamer for discussion.
This work was supported
in part
by the Centro Multidisci{\-}pli{\-}nar de Astro{\-}f\'{\i}{\-}sica (CENTRA)
and by the Funda\c{c}\~ao para a Ci\^encia e a Tecnologia (Portugal)
under grant POCTI/FNU/49529/2002.


\begin{thebibliography}{xx}
\bibitem{footnote}
Spacelike plane-wave vectors are associated with $c_{ph}<1$,
while in the current model $c_{ph}\to 1$ as
$|\vec p\hspace{.5pt}|\to\infty$ \cite{us}.
The phase-speed condition then implies an upper cutoff
for the radiated momenta and frequencies
that tends to infinity for lightlike sources.
\end{thebibliography}
\end{document}